\documentclass{elsart}

\usepackage{amssymb}

\begin{document}

\begin{frontmatter}



\title{A field-theoretical approach to the extended
Hubbard model}


\author{Z.G. Koinov}

\address{Department of Physics and Astronomy, University of Texas at San Antonio, San Antonio, TX 78249, USA}
\ead{Zlatko.Koinov@utsa.edu}
\begin{abstract}
We transform the quartic Hubbard terms in the  extended Hubbard
model to a quadratic form by making the Hubbard-Stratonovich
transformation for the electron operators. This transformation
allows us to derive exact results for mass operator and
charge-charge and spin-spin correlation functions for s-wave
superconductivity. We discuss the application of the method to the
d-wave superconductivity.
\end{abstract}
\end{frontmatter}

\section{Introduction}
The Hubbard model predicts phase instabilities which give rise to a
divergence of the charge and spin correlation functions, and
therefore, it has been the focus of particular interest as a model
for high-temperature superconductivity.  The Hamiltonian of the
standard Hubbard model contains only two terms representing the
hopping of electrons between sites of the lattice and their on-site
interaction. If the interaction between electrons on different sites
of the lattice is included, the model is referred as the extended
Hubbard model.\\ In what follows we study the following Hamiltonian:
\begin{equation}
H=-\sum_{i,j,\sigma}t_{ij}\psi^\dag_{i,\sigma}\psi_{j,\sigma}
-\mu\sum_{i,\sigma}\widehat{n}_{i,\sigma}+U\sum_i
\widehat{n}_{i,\uparrow} \widehat{n}_{i,\downarrow}
-V\sum_{<i,j>\sigma\sigma'}\widehat{n}_{i,\sigma}\widehat{n}_{j,\sigma'},
\label{Hubb1}\end{equation} where $\mu$ is the chemical potential.
The Fermi operator $\psi^\dag_{i,\sigma}$ ($\psi_{i,\sigma}$)
creates (destroys) a fermion on the lattice site $i$ with spin
projection $\sigma=\uparrow,\downarrow$ along a specified direction,
and $\widehat{n}_{i,\sigma}=\psi^\dag_{i,\sigma}\psi_{i,\sigma}$ is
the density operator on site $i$. The symbol $\sum_{<ij>}$ means sum
over nearest-neighbor sites. The first term in (\ref{Hubb1}) is the
usual kinetic energy term in a tight-binding approximation, where
$t_{ij}$ is  the single electron hopping integral. Depending on the
sign of $U$, the third term describes the on-site repulsive or
attractive interaction between electrons with opposite spins. We
assume that $V>0$, so the last term is expect to stabilize the
pairing by bringing in a nearest-neighbor attractive interaction.
The lattice spacing is assumed to be $a=1$ and  the total
number of sites is $N$.\\
The simplest method to study the possibility for the extended
Hubbard model to show a superconducting instability is to apply
mean-field analysis of pairing followed by general random phase
approximation (GRPA) \cite{KM,BR}. Going beyond the GRPA requires
reliable approximation schemes to handle self-consistent relations
between single- and two-particle quantities: the mass operator
$\Sigma$ depends on the two-particle Green function $K$, and the
kernel  of the Bethe-Salpeter (BS) equation $\delta\Sigma/\delta G$
for the spectrum of the collective excitations itself does depend on
the mass operator. A possible approximation to this problem is the
so-called two-particle self-consistent  (TPSC) approach
\cite{TPSC1}-\cite{TPSC12}. The TPSC approach is a method for
closing the set of equations for single-particle mass operator and
the two-body density matrix operator. The later can be factorized by
introducing the so-called equal-time pair-correlation function
$g_{\sigma\sigma'}(i,j)$ \cite{TPSC9,TPSC10} which itself depends on
the density-density correlation function. In other words, the TPSC
approach goes beyond the GRPA for single-particle mass operator by
establishing a self-consistency relation between single-particle and
two-particle quantities. By  setting $g_{\sigma\sigma'}(i,j)=1$ one
should recover the GRPA results for
the mass operator and charge and spin correlation functions.\\
In what follows, we first  obtain exact formulas for the electron
self-energy (electron mass operator), the charge and spin
correlation functions. We also briefly discuss how our approach
could be generalized in order to include d-wave instabilities of the
types examined in Refs. \cite{S,N,VV,AM,F}.
\section{Field-theoretical approach to extended Hubbard model}

The interaction part of the Hamiltonian (\ref{Hubb1}) is quartic in
the Grassmann fermion fields so the functional integrals cannot be
evaluated exactly. However, it is convenient to transform the
quartic Hubbard terms in (\ref{Hubb1}) to a quadratic form by making
the Hubbard-Stratonovich transformation for the electron operators:
\begin{eqnarray}&\int \mu[A]\exp\left[\widehat{\overline{\psi}}
(y)\widehat{\Gamma}^{(0)}_{\alpha}(y;x|z)\widehat{\psi}(x)A_{\alpha}(z)\right]
=\exp\{-\frac{1}{2}\widehat{\overline{\psi}}
(y)\widehat{\Gamma}^{(0)}_{\alpha}(y;x|z)\widehat{\psi}(x)
\nonumber\\&D_{\alpha,\beta}^{(0)}(z,z') \widehat{\overline{\psi}}
(y')\widehat{\Gamma}^{(0)}_{\beta}(y';x'|z')\widehat{\psi}(x')\}.\label{HSa}\end{eqnarray}
The symbol "hat" over any quantity $O$ means that this quantity is a
matrix. The functional measure $D\mu[A]$ is chosen to be:
$$
\mu[A]=DAe^{\frac{1}{2}A_{\alpha}(z)D_{\alpha,\beta}^{(0)-1}(z,z')
A_{\beta}(z')},\int \mu[A] =1.$$
 The Hubbard-Stratonovich transformation converts the quartic problem of interacting
electrons  to the more tractable quadratic problem of noninteracting
Nambu fermion  fields \begin{equation}\widehat{\overline{\psi}}
(y)=\left(\psi^\dag_\uparrow(y)
\psi_\downarrow(y) \right),\quad \widehat{\psi}(x)=\left(%
\begin{array}{c}
  \psi_\uparrow(x)  \\
  \psi^\dag_\downarrow(x) \\
\end{array}%
\right)\label{FF}\end{equation} coupled to a Bose field
$A_{\alpha}(z)$ where $\alpha=\uparrow,\downarrow$ is
 the spin degree of freedom which  reflects the spin-dependent nature of the Hubbard interaction.
The bare boson propagator in (\ref{HSa}) provides an instantaneous
spin-dependent interaction, and in accordance with the Hamiltonian
(\ref{Hubb1}), it should have the following form:
\begin{eqnarray}
&D^{(0)}_{\alpha \beta}(z,z')=D^{(0)}_{\alpha
\beta}(j,j';v-v')=\delta(v-v')\left[U\delta_{jj'}\delta_{\alpha\overline{\beta}}-2V_{<jj'>}
\left(\delta_{\overline{\alpha},\beta}+\delta_{\alpha\beta}\right)\right]
\nonumber\\&=\frac{1}{N} \sum_\textbf{k}\sum_{\omega_p}
 e^{\left\{\imath\left[\textbf{k.}\left(\textbf{r}_j-\textbf{r}_{j'}\right)
 -\omega_p\left(v-v'\right)\right]\right\}}
 D^{(0)}_{\alpha\beta}(\textbf{k};\imath\omega_p),\nonumber\\&
 D^{(0)}_{\alpha
\beta}(\textbf{k};\imath\omega_p)=
U\delta_{\overline{\alpha},\beta}-V(\textbf{k})\left(
\delta_{\overline{\alpha},\beta}+\delta_{\alpha,\beta}\right).\nonumber
\end{eqnarray}
Here $\overline{\alpha}$ is complimentary of $\alpha$, and
$V(\textbf{k})=4V\left(\cos k_x+\cos k_y\right)$ is the
nearest-neighbor interaction in momentum space. The symbol
$V_{<jj'>}$ is equal to $V$ if $j$ and $j'$ sites are nearest
neighbors, and zero otherwise. We have used composite variables
$y=\{\textbf{r}_i,u\}=\{i,u\}$,
$x=\{\textbf{r}_{i'},u'\}=\{i',u'\}$,
$z=\{\textbf{r}_{j},v\}=\{j,v\}$ and
$z'=\{\textbf{r}_{j'},v'\}=\{j',v'\}$, where
$\textbf{r}_{i},\textbf{r}_{i'},\textbf{r}_{j}$ and
$\textbf{r}_{j'}$
 are the lattice site vectors. The symbol $\sum_{\omega_p}$ is used
to denote $\beta^{-1}\sum_{p}$. For boson fields we have
$\omega_{p}=(2\pi/ \beta)p ;
p=0, \pm 1, \pm 2,...$.\\
After performing the Hubbard-Stratonovich transformation, the
 action of the system becomes
 \begin{equation}S= S^{(e)}_0+S^{(A)}_0+S^{(e-A)},\label{Ac}\end{equation}
where:
\begin{equation}
S^{(e)}_0=\widehat{\overline{\psi}
}(y)\widehat{G}^{(0)-1}(y;x)\widehat{\psi} (x),\label{GelAct}
\end{equation}
\begin{equation}
S^{(A)}_0=\frac{1}{2}A_{\alpha}(z)D^{(0)-1}_{\alpha
\beta}(z,z')A_{\beta}(z'),\label{GphAct}
\end{equation}
\begin{equation}
S^{(e-A)}=\widehat{\overline{\psi}}
(y)\widehat{\Gamma}^{(0)}_{\alpha}(y,x\mid z)\widehat{\psi}
(x)A_{\alpha}(z).\label{Gelph}
\end{equation}
 The inverse Green function  of free electrons
$\widehat{G}^{(0)-1}(y;x)$ is diagonal with respect to the spin
indices and has its usual form:
\begin{eqnarray}
&\widehat{G}^{(0)-1}(y;x)=\left(%
\begin{array}{cc}
  G^{(0)-1}(\uparrow,y;\uparrow,x) & 0  \\
  0 & -G^{(0)-1}(\downarrow,y;\downarrow,x) \\
\end{array}%
\right)\nonumber\\&=\frac{1}{N}\sum_\textbf{k}\sum_{\omega_m}\exp\{\imath
\textbf{k.}(\textbf{r}_i-\textbf{r}_{i'}-\omega_m(u-u')\}\times\nonumber\\&
\left(%
\begin{array}{cc}
  G_{\uparrow,\uparrow}^{(0)-1}(\textbf{k},\imath\omega_m) & 0  \\
  0 & -G_{\downarrow,\downarrow}^{(0)-1}(\textbf{k},\imath\omega_m) \\
\end{array}%
\right), \label{EGF0}
\end{eqnarray}
 where $G_{\uparrow,\uparrow}^{(0)-1}(\textbf{k},\imath\omega_m)=\left[\imath\omega_m-
\left(\epsilon(\textbf{k})-\mu\right)\right]^{-1},$ and $
G_{\downarrow,\downarrow}^{(0)-1}(\textbf{k},\imath\omega_m)=\left[\imath\omega_m+
\left(\epsilon(\textbf{k})-\mu\right)\right]^{-1}$. Here
$\epsilon(\textbf{k})=-2t(\cos k_x+\cos k_y)-4t'\cos k_x\cos k_y$ is
the non-interacting dispersion on a square lattice, $\mu$ is the
electron chemical potential, and the symbol $\sum_{\omega_m}$ is
used to denote $\beta^{-1}\sum_{m}$. For fermion fields we have
$\omega_m=
   (2\pi/\beta)(m +1/2) ;m=0, ±1, ±2,… $.\\
 The bare vertex
$\widehat{\Gamma}^{(0)}_{\alpha}(y;x\mid z)$ is a $2\times 2$ matrix
defined as follows:
\begin{eqnarray}
&\widehat{\Gamma}^{(0)}_{\alpha}(y;x\mid
z)=\left(%
\begin{array}{cc}
  \Gamma^{(0)}_{\alpha}(\uparrow,y;\uparrow x\mid
z) & 0  \\
  0 & -\Gamma^{(0)}_{\alpha}(\downarrow,y;\downarrow,x\mid
z) \\
\end{array}%
\right), \nonumber\\& \Gamma^{(0)}_{\alpha}(\sigma,y;\sigma,x\mid
z)=\Gamma^{(0)}_{\alpha}(\sigma,i,u;\sigma,i',u'\mid
i'',v)\\&=\delta(u-v)\delta(u-u')\delta_{\sigma,\alpha}\delta_{i,i'}\delta_{i,i''}
.\nonumber
\end{eqnarray}
Since
 the electrons polarize the boson field, and the boson field acts onto the electrons,
 our approach describes the
  correlated motion of the electrons and the surrounding polarization
  field.\\
 In field theory the expectation value of a general operator
$\widehat{O}(u)$ is expressed as a functional integral over the
boson field $A$ and the Grassmann fermion fields
$\widehat{\overline{\psi}}$ and $\widehat{\psi}$
\begin{eqnarray}&<\widehat{T}_u(\widehat{O}(u))>=\frac{1}{Z[J,M]}\int
D\mu[\widehat{\overline{\psi}},\widehat{\psi},A]\widehat{O}(u)\times\nonumber\\&
\exp\left[J_{\alpha}(z)A_{\alpha}(z)-\widehat{\overline{\psi}}
(y)\widehat{M}(y;x)\widehat{\psi}
(x)\right]|_{J=M=0},\label{Ex}\end{eqnarray} where the symbol
$<...>$ means that the thermodynamic average is made, and
$\widehat{T}_u$ is an $u-$ordering operator. $J, M$ are the sources
of the boson and fermion fields, respectively. The functional
$Z[J,M]$ is defined by
\begin{equation}
Z[J,M]=\int D\mu[\widehat{\overline{\psi}},\widehat{\psi},A]
e^{[J_{\alpha}(z)A_{\alpha}(z)-\widehat{\overline{\psi}}
(y)\widehat{M}(y;x)\widehat{\psi} (x)]},\label{GW}
\end{equation}where the functional measure
$D\mu[\widehat{\overline{\psi}},\widehat{\psi},A]=DAD\widehat{\overline{\psi}}D\widehat{\psi}
\exp\left(S\right)$ satisfies the condition $\int
D\mu[\widehat{\overline{\psi}},\widehat{\psi},A]=1$.\\
It is convenient to introduce complex indices $1=\{\sigma_1,x_1\}$,
$2=\{\sigma_2,y_2\},...$, where,
$\sigma_{1,2}=\{\uparrow,\downarrow\}$ and
$x_1=\{\textbf{r}_{i_1},u_1\}$, and $y_2=\{\textbf{r}_{i_2},u_2\}$.
We define a functional derivative $\delta / \delta M(1;2)$, and
depending on the spin degrees of freedom  $\sigma_1$ and $\sigma_2$,
there are four possible derivatives:$$\frac{\delta}{\delta
M(\uparrow,y_2;\uparrow,x_1)},\quad \frac{\delta}{\delta
M(\uparrow,y_2;\downarrow,x_1)},$$$$ \frac{\delta}{\delta
M(\downarrow,y_2;\uparrow,x_1)},\quad \frac{\delta}{\delta
M(\downarrow,y_2;\downarrow,x_1)}.$$ The reason to write the
expectation value (\ref{Ex}) as a functional integral is that all
Green functions related to system under consideration can be
expressed in terms of the functional derivatives of the generating
functional of the connected Green functions $W[J,M]=\ln Z[J,M]$. By
means of the functional $W[J,M]$, we define the following Green and
vertex
functions of the extended Hubbard model:\\
 Boson Green function:
\begin{equation}D_{\alpha \beta}(z,z')=-\frac{\delta^2W}{\delta
J_{\alpha}(z)\delta J_{\beta}(z')}; \label{PGF}
\end{equation}
The single-electron Green function $G(1;2)=-\delta W/\delta M(2;1)$
in the Hubbard model assumes the form:
\begin{equation}\widehat{G}(1;2)= -\left(%
\begin{array}{cc}
  <\widehat{T}_u\left(\psi_\uparrow(x_1)\psi^\dag_\uparrow(y_2)\right)> &
   <\widehat{T}_u\left(\psi_\uparrow(x_1)\psi_\downarrow(y_2)\right)>    \\
  <\widehat{T}_u\left(\psi^{\dag}_\downarrow(x_1)\psi^\dag_\uparrow(y_2)\right)> & <\widehat{T}_u\left(\psi^\dag_\downarrow(x_1)\psi_\downarrow(y_2)\right)> \\
\end{array}%
\right). \label{EGF}
\end{equation}
Depending on the two spin degrees of freedom $\sigma_1$ and
$\sigma_2$, there exist two "normal" Green functions
$$G(\uparrow, x_1;\uparrow, y_2)=G_{\uparrow,\uparrow}( x_1, y_2)=-<\widehat{T}_u\left(\psi_\uparrow(x_1)\psi^\dag_\uparrow(y_2)\right)>=
-\frac{\delta W}{\delta M(\uparrow, y_2;\uparrow, x_1)},$$
$$G(\downarrow, x_1;\downarrow, y_2)=G_{\downarrow,\downarrow}( x_1, y_2)=-<\widehat{T}_u\left(\psi^\dag_\downarrow(x_1)\psi_\downarrow(y_2)
\right)>= -\frac{\delta W}{\delta M(\downarrow, y_2;\downarrow,
x_1)},$$ and two "anomalous" Green functions
$$G(\downarrow, x_1;\uparrow, y_2)=G_{\downarrow,\uparrow}( x_1,y_2)=-<\widehat{T}_u\left(\psi^\dag_\downarrow(x_1)
\psi^\dag_\uparrow(y_2)\right)>= -\frac{\delta W}{\delta M(\uparrow,
y_2;\downarrow, x_1)},$$
$$G(\uparrow, x_1;\downarrow, y_2)=G_{\uparrow,\downarrow}( x_1, y_2)=-<\widehat{T}_u\left(\psi_\uparrow(x_1)
\psi_\downarrow(y_2)\right)>= -\frac{\delta W}{\delta M(\downarrow,
y_2;\uparrow, x_1)}.$$We introduce Fourier transforms of the
"normal" $G_{\uparrow,\uparrow}(\textbf{k},u-u')=
-<\widehat{T}_u(\psi_{\uparrow,\textbf{k}}(u')\psi^\dag_{\uparrow,\textbf{k}}(u))>$,
$G_{\downarrow,\downarrow}(\textbf{k},u-u')=-<\widehat{T}_u(\psi^\dag_{\downarrow,\textbf{k}}(u')
\psi_{\downarrow,\textbf{k}}(u))>$ and "anomalous"
$G_{\downarrow,\uparrow}(\textbf{k},u-u')=-<\widehat{T}_u(\psi_{\uparrow,\textbf{k}}(u')
\psi_{\downarrow,\textbf{k}}(u))>$,
$G_{\uparrow,\downarrow}(\textbf{k},u-u')=-<\widehat{T}_u(\psi^\dag_{\downarrow,\textbf{k}}(u')
\psi^\dag_{\uparrow,\textbf{k}}(u))>$ one-particle Green functions.
Here $\psi^+_{\uparrow,\textbf{k}}(u),\psi_{\uparrow,\textbf{k}}(u)$
and
$\psi^+_{\downarrow,\textbf{k}}(u),\psi_{\downarrow,\textbf{k}}(u)$
are the creation-annihilation Heisenberg operators. The final form
of  single-particle Green function is given by
\begin{eqnarray}&
\widehat{G}(1;2)=
\frac{1}{N}\sum_{\textbf{k}}\sum_{\omega_{m}}\exp\{\imath\left[\textbf{k.}\left(
\textbf{r}_{i_1}-\textbf{r}_{i_2}\right)-\omega_{m}(u_1-u_2)\right]\}
\\&\left(%
\begin{array}{cc}
  G_{\uparrow,\uparrow}(\textbf{k},\imath\omega_m) & G_{\uparrow,\downarrow}(\textbf{k},\imath\omega_m)  \\
 G_{\downarrow,\uparrow}(\textbf{k},\imath\omega_m) & G_{\downarrow,\downarrow}(\textbf{k},\imath\omega_m) \\
\end{array}%
\right).\nonumber\end{eqnarray}
  The two-particle
Green function is defined by:
\begin{equation}
K\left(%
\begin{array}{cc}
  1 & 3  \\
  2 & 4 \\
\end{array}%
\right)=\frac{\delta^2 W}{\delta M(2;1)\delta M(3;4)}=-\frac{\delta
G(1;2)}{\delta M(3;4)} ; \label{TGF}
\end{equation}
Depending on the four spin degrees of freedom
$\sigma_1,\sigma_2,\sigma_3$ and $\sigma_4$, there are sixteen
different components of the two-particle Green
function.\\
 The vertex function $\Gamma_{\sigma_\alpha}(2;1 \mid z)$ is given by:
\begin{eqnarray}&
\Gamma_{\alpha}(2;1 \mid z)=-\frac{\delta G^{-1}(2;1)}{\delta
J_{\beta}(z')} D^{-1}_{\beta
\alpha}(z',z)\nonumber\\&=G^{-1}(2;3)\frac{\delta G(3;4)}{\delta
J_{\beta}(z')}G^{-1}(4;1) D^{-1}_{\beta \alpha}(z',z) .
\label{VF}\end{eqnarray} If the spin variable $\alpha$ is fixed,
then depending on the spin degrees of freedom $\sigma_1$ and
$\sigma_2$, there are four different vertex functions. \\The action
(\ref{Ac}) as well as all of the above definitions allow us to map
the extended Hubbard model onto the polariton model that describes
the light propagation in semiconductors \cite{ZK}. This mapping
allows us to apply directly to the Hubbard model all exact equations
and relationships derived for the case of light propagation in
crystals. For example, we can write the mass operator
$\Sigma(1;2)=\Sigma_H(1;2)+\Sigma_F(1;2)$ as a sum of Hartree and
Fock parts \cite{ZK}. The Hartree contribution to the mass operator
is diagonal with respect to the spin indices:
\begin{eqnarray}
&\Sigma_H(1;2)=\Sigma_H(\sigma_1,i_1,u_1;\sigma_2,i_2,u_2)\nonumber\\&=
\Gamma^{(0)}_{\alpha}(1;2\mid
z)G(4;3)\Gamma^{(0)}_{\beta}(3;4\mid z')D^{(0)}_{\alpha\beta
}(z,z')=\delta(1-2)\times\nonumber\\&\{-U
G_{\overline{\sigma}_1,\overline{\sigma}_1}(i_1,u_1;i_1,u_1)\nonumber\\&
-2V\sum_{a}[ G_{\sigma_1,\sigma_1}(i_1,u_1;i_1+a,u_1)-
G_{\overline{\sigma
}_1,\overline{\sigma}_1}(i_1,u_1;i_1+a,u_1)]\},\label{H}\end{eqnarray}
where the summation on $a$ runs over the nearest-neighbor sites of
site $i_1$.\\ The Fock part depends on the boson Green function $D$,
or equivalently, depends on the two-particle Green function $K$:
\begin{eqnarray}
&\Sigma_F(1;2)=\Sigma_F(\sigma_1,i_1,u_1;\sigma_2,i_2,u_2)\nonumber\\&
=-\Gamma^{(0)}_{\alpha}(1;3\mid z)G(3;4)\Gamma_{\beta}(4;2\mid z')
D_{\alpha \beta}(z,z'),\nonumber\\&=-\Gamma^{(0)}_{\alpha}(1;6|z)
D^{(0)}_{\alpha\beta}(z,z')\Gamma^{(0)}_{\beta}(4;5|z')
K\left(%
\begin{array}{cc}
  5 & 3  \\
  4 & 6 \\
\end{array}%
\right)G^{-1}(3;2)\nonumber\\&=\{U
K\left(%
\begin{array}{cc}
  \overline{\sigma}_1,i_1,u_1 & \sigma_3,i_3,u_3  \\
  \overline{\sigma}_1,i_1,u_1 & \sigma_1,i_1,u_1 \\
\end{array}%
\right)+2V\sum_{a}[K\left(%
\begin{array}{cc}
  \sigma_1,i_1+a,u_1 & \sigma_3,i_3,u_3  \\
  \sigma_1,i_1+a,u_1 & \sigma_1,i_1,u_1 \\
\end{array}%
\right)\nonumber\\&-K\left(%
\begin{array}{cc}
  \overline{\sigma}_1,i_1+a,u_1 & \sigma_3,i_3,u_3  \\
  \overline{\sigma}_1,i_1+a,u_1 & \sigma_1,i_1,u_1 \\
\end{array}%
\right)]\}G^{-1}(\sigma_3,i_3,u_3
;\sigma_2,i_2,u_2).\label{FT}\end{eqnarray} The $V$-terms in
(\ref{FT}) differ from the corresponding result in
\cite{TPSC9,TPSC10}.
\section{Spectrum of the collective modes}
 Spectrum
of the two-particle excitations (or collective modes)
$\omega(\textbf{Q})$ can be obtained by locating the positions of
the common poles of the Fourier transform of the two-particle
fermion Green function (\ref{TGF}) and the Fourier transform of the
boson Green function (\ref{PGF}). In other words, collective modes
are defined by the solutions of the BS equation for the function
$K$ or the Dyson equation for the boson  function $D$.\\
The BS equation is $K^{-1}\Psi=\left[K^{(0)-1}-I\right]\Psi=0$,
where $I$ is the kernel, and $$K^{(0)-1}\left(%
\begin{array}{cc}
  1 & 3  \\
  2 & 4 \\
\end{array}%
\right)=G^{-1}(1;3)G^{-1}(4;2)$$ is the two-particle free propagator
constructed from a pair of fully dressed single-particle Green
functions. The kernel $I=\frac{\delta\Sigma_H}{\delta
G}+\frac{\delta\Sigma_F}{\delta G}$ depends on the functional
derivative of the Fock contribution to the mass operator. Since the
Fock term itself depends on the two-particle Green function $K$ (see
Eq. (\ref{FT})), we have to solve self-consistently a set of two
equations, namely the BS equation and the Dyson equation
$G^{-1}=G^{(0)-1}-\Sigma$ for the single-particle Green function.
\\
Similar obstruction arises if we start from the Dyson equation for
boson Green function:
\begin{equation}D^{-1}_{\alpha \beta}(z,z') =D^{(0)-1}_{\alpha
\beta}(z,z')-\widetilde{\Pi}_{\alpha
\beta}(z,z'),\label{DE}\end{equation}
 where
 $\widetilde{\Pi}_{\alpha
\beta}(z,z') $ is the proper self-energy of the boson field.
 The obstruction now is that
the proper self-energy depends on the vertex function $\Gamma$, or
on the two-particle Green function $\widetilde{K}$:
\begin{eqnarray}&
\widetilde{\Pi}_{\alpha \beta}(z,z')=\Gamma^{(0)}_{\alpha}(1;2\mid
z)\widetilde{K}\left(%
\begin{array}{cc}
  2 & 3  \\
  1 & 4 \\
\end{array}%
\right)\Gamma^{(0)}_{\beta}(3;4\mid z')\nonumber\\&=
\Gamma^{(0)}_{\alpha}(1;2\mid z)G(2;3)G(4;1) \Gamma_{\beta}(3;4\mid
z'). \label{P}\end{eqnarray} The Green function $\widetilde{K}$,
which we shall call the Green
 function of electronic excitations,
satisfies the following BS equation:
\begin{equation}
 \widetilde{K}^{-1}\left(%
\begin{array}{cc}
  1 & 3  \\
  2 & 4 \\
\end{array}%
\right)= K^{(0)-1}\left(%
\begin{array}{cc}
  1 & 3  \\
  2 & 4 \\
\end{array}%
\right)-\frac{\delta \Sigma_F(1;2)}{\delta G(3;4)}. \label{GFGME}
\end{equation}
Let us introduce the so-called general response function
$\Pi_{\alpha\beta}$:
\begin{equation}\Pi_{\alpha\beta}(z;z')=\Gamma^{(0)}_{\alpha}(1;2\mid
z)K\left(%
\begin{array}{cc}
  2 & 3  \\
  1 & 4 \\
\end{array}%
\right)\Gamma^{(0)}_{\beta}(3;4\mid z').\label{KM2}\end{equation} By
means of (\ref{KM2}), we can rewrite the BS and Dyson equations as
follows: $K=K^{(0)}+K^{(0)}\Gamma^{(0)}_\alpha
\Pi_{\alpha\beta}\Gamma^{(0)}_\beta K^{(0)}$, $D=D^{(0)}+D^{(0)}\Pi
D^{(0)}$. The Fourier transforms of the general response function
$\Pi_{\alpha\beta}(\textbf{Q};\omega)$ and the proper self-energy
$\widetilde{\Pi}_{\alpha\beta}(\textbf{Q};\omega)$ are connected by
the following equation:
\begin{eqnarray}&\Pi_{\alpha\beta}(\textbf{Q};\omega)=\widetilde{\Pi}_{\alpha\beta}(\textbf{Q};\omega)+
\widetilde{\Pi}_{\alpha\gamma}(\textbf{Q};\omega)\left(-U+V(\textbf{Q})
\right)\Pi_{\overline{\gamma}\beta}(\textbf{Q};\omega)\nonumber\\&+
\Pi_{\alpha\gamma}(\textbf{Q};\omega)V(\textbf{Q})
\widetilde{\Pi}_{\gamma\beta}(\textbf{Q};\omega),\label{KM1}\end{eqnarray}
 According to Eq. (\ref{P}), the proper self-energy $\widetilde{\Pi}$,
the Green function $\widetilde{K}$ and the vertex function $\Gamma$
must have common poles. Let $E_{l}(\textbf{Q})$ and $\textbf{Q}$
denote the energy and momentum of these common poles. Close to
$E_{l}(\textbf{Q})$ one can write:
\begin{eqnarray}
&\widetilde{K}\left(%
\begin{array}{cc}
  1 & 3  \\
  2 & 4 \\
\end{array}%
\right)=\widetilde{K}\left(%
\begin{array}{cc}
  \sigma_1,i_1,u_1 & \sigma_3,i_3,u_3  \\
  \sigma_2,i_2,u_2 & \sigma_4,i_4,u_4 \\
\end{array}%
\right)\nonumber\\&\approx \sum_{\omega_{p}}e^{-\imath
\omega_{p}(u_1-u_3)}\frac{\Phi_{\sigma_1,\sigma_2}^{l\textbf{Q}}(\textbf{r}_{i_2},\textbf{r}_{i_1};u_2-u_1)
\Phi_{\sigma_3,\sigma_4}^{l\textbf{Q} \ast
}(\textbf{r}_{i_3},\textbf{r}_{i_4};u_4-u_3)}{\imath \omega_p -
\omega_{l\textbf{Q}}},\label{Gf}\end{eqnarray} where
$\Phi_{\sigma_1,\sigma_2}^{l\textbf{Q}}(\textbf{r}_{i_2},\textbf{r}_{i_1};u_1-u_2)$
are the BS amplitudes:
$$\Phi_{\sigma_1,\sigma_2}^{l\textbf{Q}}(\textbf{r}_{i_2},\textbf{r}_{i_1};u_1-u_2)=\exp
\left[\imath
\frac{1}{2}\textbf{Q.}\left(\textbf{r}_{i_1}+\textbf{r}_{i_2}\right)\right]
\phi^{l\textbf{Q}}_{\sigma_1,\sigma_2}(\textbf{r}_{i_1}-\textbf{r}_{i_2};u_1-u_2).$$
Due to the form of the bare vertex $\widehat{\Gamma}^{(0)}$, we have
to take into account only the equal "time" $u_1=u_2$ amplitudes:
$$\phi^{l\textbf{Q}}_{\sigma_1,\sigma_2}(\textbf{r}_{i_1}-\textbf{r}_{i_2};0)=
\frac{1}{N}\sum_{\textbf{k}} \exp\{\imath\textbf{k.}
\left(\textbf{r}_{i_1}-\textbf{r}_{i_2}\right)\}\phi^l_{\sigma_1,\sigma_2}(\textbf{k},\textbf{Q}),$$
where $\phi^l_{\downarrow\uparrow}(\textbf{k},\textbf{Q}),
\phi^l_{\uparrow\downarrow}(\textbf{k},\textbf{Q}),\phi^l_{\uparrow\uparrow}(\textbf{k},\textbf{Q})$
and $\phi^l_{\downarrow\downarrow}(\textbf{k},\textbf{Q})$ are the
equal "time" two-particle wave functions in
$\textbf{k}$-representation.  By means of (\ref{Gf}) and (\ref{P})
we obtain:
\begin{equation}
\widetilde{\Pi}_{\alpha\beta}(\textbf{Q},\omega)=\sum_{l}\left[
\frac{\phi^{l\textbf{Q}}_{\alpha,\alpha}(0;0)\phi^{l\textbf{Q}*}_{\beta,\beta}(0;0)}{\omega-\omega_{l\textbf{Q}}+\imath
0^+}
-\frac{\phi^{l\textbf{Q}*}_{\alpha,\alpha}(0;0)\phi^{l\textbf{Q}}_{\beta,\beta}(0;0)}{\omega+\omega_{l\textbf{Q}}+\imath
0^+} \right],\label{PO}
\end{equation}
where $\omega_{l\textbf{Q}}=E_{l}(\textbf{Q})-\mu$. If the proper
self-energy is known, the spectrum of the collective excitations
$\omega(\textbf{Q})$ could be obtained assuming the vanishing of the
following $2\times 2$ determinant:
\begin{equation}
det\parallel \delta_{\alpha,\beta}-\left(-U+V(\textbf{Q})
\right)\widetilde{\Pi}_{\overline{\alpha}\beta}(\textbf{Q},\omega)+V(\textbf{Q})
\widetilde{\Pi}_{\alpha\beta}(\textbf{Q},\omega)\parallel=0.\label{PSp}
\end{equation}
 By solving Eq. (\ref{PSp}) (with the help of
$\widetilde{\Pi}_{\alpha\beta}=\widetilde{\Pi}_{\beta\alpha}$) we
find two different types of collective modes. The first one is
governed by $U$ interaction:
\begin{equation}
0=1-U\left[\widetilde{\Pi}_{\uparrow\uparrow}(\textbf{Q},\omega)-
\widetilde{\Pi}_{\uparrow\downarrow}(\textbf{Q},\omega)\right].
\label{EQ1}\end{equation} Strictly speaking the $V$ interaction is
included indirectly in Eq. (\ref{EQ1}) through BS amplitudes and
poles of the function $\widetilde{K}$. This collective mode
manifests itself as a pole of the spin response function
$\chi_{ss}(\textbf{Q};\omega)=2\left[\Pi_{\uparrow\uparrow}(\textbf{Q};\omega)-
\Pi_{\uparrow\downarrow}(\textbf{Q};\omega)\right]$, and therefore,
gives rise to the spin instabilities. The spin correlation function
is defined by
\begin{equation}
\chi_{ss}(\textbf{Q};\omega)=\frac{\widetilde{\chi}_{ss}(\textbf{Q};\omega)}
{1-\widetilde{\chi}_{ss}(\textbf{Q};\omega)U/2},
\label{SS}\end{equation} where
$\widetilde{\chi}_{ss}(\textbf{Q};\omega)=2\left[\widetilde{\Pi}_{\uparrow\uparrow}
(\textbf{Q};\omega)-
\widetilde{\Pi}_{\uparrow\downarrow}(\textbf{Q};\omega)\right]$.\\
The second type of collective modes satisfies the following
equation:
\begin{equation}
0=1+\left[U-2V(\textbf{Q})\right]\left[\widetilde{\Pi}_{\uparrow\uparrow}(\textbf{Q},\omega)
+\widetilde{\Pi}_{\uparrow\downarrow}(\textbf{Q},\omega)\right].
\label{EQ2}\end{equation}
 The second collective mode
manifests itself as a pole of the charge response function
$\chi_{cc}(\textbf{Q};\omega)=2\left[\Pi_{\uparrow\uparrow}(\textbf{Q};\omega)+
\Pi_{\uparrow\downarrow}(\textbf{Q};\omega)\right]$ which is
determined by the following equations:
\begin{equation}
\chi_{cc}(\textbf{Q};\omega)=\frac{\widetilde{\chi}_{cc}(\textbf{Q};\omega)}
{1+\widetilde{\chi}_{cc}(\textbf{Q};\omega)\left[U/2-V(\textbf{Q})\right]},
\label{CC}\end{equation} where
$\widetilde{\chi}_{cc}(\textbf{Q};\omega)=2\left[\widetilde{\Pi}_{\uparrow\uparrow}(\textbf{Q};\omega)+
\widetilde{\Pi}_{\uparrow\downarrow}(\textbf{Q};\omega)\right]$.\\
Within the GRPA, one should replace $\widetilde{K}$ in (\ref{P}) by
$K^{(0)}$, thus obtaining the free response functions $\chi^{(0)}$
instead of the exact $\widetilde{\chi}$ expressions. In other words,
the exact relations (\ref{SS}) and (\ref{CC}) in the GRPA are given
by
\begin{equation}
\chi_{ss}(\textbf{Q};\omega)=\frac{\chi^{(0)}_{ss}(\textbf{Q};\omega)}
{1-\chi^{(0)}_{ss}(\textbf{Q};\omega)U/2}, \label{SS1}\end{equation}
\begin{equation}
\chi_{cc}(\textbf{Q};\omega)=\frac{\chi^{(0)}_{cc}(\textbf{Q};\omega)}
{1+\chi^{(0)}_{cc}(\textbf{Q};\omega)\left[U/2-V(\textbf{Q})\right]}.
\label{CC1}\end{equation}

\section{Discussion }
We have established exact results for mass operator and spin and
charge correlation functions in the case of s-wave superconductivity
described by the extended Hubbard Hamiltonian. The Hamiltonian
depends on an on-site repulsive interactions $U$, which drives
antiferromagnetism, and a near-neighbor attractive interactions V,
which drives d-wave superconductivity. Let us discuss consider the
extended Hubbard model with a nearest-neighbor repulsion $V$. In
this case the near-neighbor repulsive interaction V drives
instabilities related to the change in Fermi surface topology
(Pomeranchuk instability) \cite{VV,AM}. The Fermi surface is related
to the diagonal elements of matrix (\ref{FT}). In the RPA we
replace  $K\left(%
\begin{array}{cc}
  1 & 3  \\
  2 & 4
\end{array}%
\right)$  by $ G(1;3)G(4;2)$, and therefore, the $U$ term in
(\ref{FT}) does not create any changes in the non-interacting Fermi
surface $\mu=\epsilon(\textbf{k})$. In RPA only the near-neighbor
repulsive interaction (V$\rightarrow$ -V) changes the Fermi surface
topology through the diagonal elements of the mass operator
(\ref{FT}):
\begin{equation}\left(\Sigma_F(\textbf{k})\right)_{\sigma\sigma}=-4V\sum_\textbf{q}
\left[\cos(k_x-q_x)+\cos(k_y-q_y)\right]n_{\sigma}(\textbf{q}),\label{VV}\end{equation}
where
$n_{\sigma}(\textbf{k})=\sum_{\omega_m}G_{\sigma\sigma}(\textbf{k},\imath
\omega_m)$ is the occupation of the site $\textbf{k}$ in momentum
space. Note that the $V$-terms of $\Sigma_F(1,2)$ in Refs.
\cite{TPSC9,TPSC10} are proportional to the delta function
$\delta(1-2)$, and therefore, $\Sigma_F(\textbf{k})$ does not depend
on $\textbf{k}$.

 It is found \cite{VV} that the near-neighbor
repulsive interaction V enhance small anisotropies producing
deformations of the Fermi surface that break the point group
symmetry of the square lattice at the Van Hove filling. Since the
antiferromagnetic order is favored by U but suppressed by V, one has
to expect that the Pomeranchuk instability competes with magnetic
instabilities and will be suppressed at some
critical value V.\\
The analog of the BCS reduced Hamiltonian for singlet density-wave
order is the so-called f-Hamiltonian \cite{N,AM,F}:
\begin{equation}
H
=\sum_{\textbf{k},\sigma}(\epsilon(\textbf{k})-\mu)n_{\textbf{k},\sigma}
+\frac{1}{2V}\sum_{\textbf{k},\textbf{p},\textbf{q}}S(\textbf{q})d_\textbf{k}d_\textbf{p}
n_{\textbf{k}}(\textbf{q})n_{\textbf{p}}(\textbf{-q}),\label{MH}\end{equation}
where $d_\textbf{k}=\cos(k_x)-\cos(k_y)$
 and the  operator
$n_{\textbf{k}}(\textbf{q})$ is
$n_{\textbf{k}}(\textbf{q})=\sum_\sigma
\psi^\dag_\sigma(\textbf{k}+\frac{1}{2}\textbf{q})\psi_\sigma(\textbf{k}-\frac{1}{2}\textbf{q})
$. In momentum space the corresponding order parameter can be
defined through the following operator (for more general definition
of $O(\textbf{q})$, see Ref. \cite{F})
\begin{equation}
O(\textbf{q})=\sum_{\textbf{k}}d_\textbf{k}
\widehat{\psi}^\dag(\textbf{k}+\frac{1}{2}\textbf{q})\widehat{\sigma}_z
\widehat{\psi}(\textbf{k}-\frac{1}{2}\textbf{q}),
\label{dO}\end{equation} where
$\widehat{\overline{\psi}}(\textbf{k})$ and
$\widehat{\psi}(\textbf{k})$ are the Fourier transforms of the Numbu
fermion fields, and $\widehat{\sigma}_z$ is the Pauli matrix. The
interaction term in the above Hamiltonian in coordinate space is
$\frac{1}{2}\sum_{i,j}S(\textbf{r}_i-\textbf{r}_j)O(\textbf{r}_i)O(\textbf{r}_j)$,
where the Fourier transform of $S(\textbf{r}_i)$ is $S(\textbf{k})$.
Now, we introduce a boson field $B(z)=B(\textbf{r}_j;v)$ with a free
propagator
$S^{(0)}(z,z')=\delta(v-v')S^{(0)}(\textbf{r}_j-\textbf{r}_{j'})$
and a bare vertex function
$\Gamma^{(0)}(y,x|z)=\Gamma^{(0)}(\textbf{r}_i,u;\textbf{r}_{i'},u'|\textbf{r}_j,v)=\delta(u-u')
\delta(u-v)\delta(\textbf{r}_i-\textbf{r}_{i'})
\delta(\textbf{r}_i-\textbf{r}_j)$. Now, we can perform exactly the
same steps as in the case of s-wave superconductivity. As a result,
one can obtain the following equation for d-wave response function
$\Pi(\textbf{Q},\omega)$:
\begin{equation}\Pi(\textbf{Q},\omega)=\widetilde{\Pi}(\textbf{Q},\omega)+
\widetilde{\Pi}(\textbf{Q},\omega)S^{(0)}(\textbf{Q})\Pi(\textbf{Q},\omega),\label{dRF}\end{equation}
where the proper self-energy is defined as
$$\widetilde{\Pi}(z,z')=\Gamma^{(0)}(1;2|z)G(2;3)G(3;4)\Gamma(3;4|z).$$
Solving (\ref{dRF}) we obtain the GRPA result for the d-wave
response function:
$$\chi(\textbf{Q},\omega)=\frac{\chi^{(0)}(\textbf{Q},\omega)}
{1-S^{(0)}(\textbf{Q})\chi^{(0)}(\textbf{Q},\omega)},$$ where
$\chi^{(0)}$ is the d-wave proper self-energy in RPA (see eq. (7) in
Ref. \cite{AM}).

   In summary, we have used the Hubbard-Stratonovich transformation to
convert the quartic Hubbard problem of interacting electrons into
more tractable quadratic problem of noninteracting electrons coupled
to a Bose field. This field-theoretical approach allows us to
express all quantities of interest in terms of the corresponding
Green functions and obtain exact relations between single- and
two-particle quantities.

\end{document}